\documentclass[
  aps,              
  pra,              
 reprint,          
  showkeys,         
  groupedaddress,   
superscriptaddress, 
  ]{revtex4-2}

\usepackage{amsmath}
\usepackage{amssymb}
\usepackage{fourier}

\usepackage{graphicx,dcolumn,bm,algpseudocode,mathtools,cases,float}

\usepackage[utf8]{inputenc}
\usepackage[T1]{fontenc}

\usepackage[hidelinks]{hyperref}
\hypersetup{
  colorlinks=true,
  allcolors=blue,
  urlcolor=blue
  }

\usepackage[dvipsnames]{xcolor}
\usepackage{totcount}
\newcounter{tofixn}

\regtotcounter{tofixn}

\begin{document}

\title[Zhang, et al.: Capillary waves from ultrasound]{Onset of low-frequency capillary waves driven by high-frequency ultrasound}

\author{Shuai Zhang}
\affiliation{Medically Advanced Devices Laboratory, Center for Medical Devices,}
\affiliation{Department of Materials Science and Engineering,}
\affiliation{Department of Mechanical and Aerospace Engineering,\\
    Jacobs School of Engineering\\
    University of California San Diego,
    La Jolla, CA 92093-0411 USA
    }
\author{Jeremy Orosco}
\affiliation{Medically Advanced Devices Laboratory, Center for Medical Devices,}
\affiliation{Department of Mechanical and Aerospace Engineering,\\
    Jacobs School of Engineering\\
    University of California San Diego,
    La Jolla, CA 92093-0411 USA
    }
\author{James Friend}
\email[Corresponding author: ]{jfriend@ucsd.edu}
\affiliation{Medically Advanced Devices Laboratory, Center for Medical Devices,}
\affiliation{Department of Materials Science and Engineering,}
\affiliation{Department of Mechanical and Aerospace Engineering,\\
    Jacobs School of Engineering\\
    University of California San Diego,
    La Jolla, CA 92093-0411 USA
    }

\date{\today}

\begin{abstract}

High frequency thickness mode ultrasound is an energy-efficient way to atomize high-viscosity fluid at high flow rate into fine aerosol mists of micron-sized droplet distributions. However the complex physics of the atomization process is not well understood. It is found that with low power the droplet vibrates at low frequency ($\mathcal{O}(10^{2})\text{ Hz}$) when driven by high-frequency ultrasound ($\mathcal{O}(10^{6})\text{ Hz}$ and above). To study the mechanism of the energy transfer that spans these vastly different timescales, we measure the droplet's interfacial response to 6.6~MHz ultrasound excitation using high-speed digital holography---a revolutionary method for capturing three dimensional surface dynamics at nanometer space and microsecond time resolutions. We show that the onset of low-frequency capillary waves is driven by feedback interplay between the acoustic radiation pressure distribution and the droplet surface. These dynamics are mediated by the Young-Laplace boundary between the droplet interior and ambient environment. Numerical simulations are performed via global optimization against the rigorously defined interfacial physics. The proposed pressure-interface feedback model is explicitly based on the pressure distribution hypothesis. For low power acoustic excitation, the simulations reveal a stable oscillatory feedback that induces capillary wave formation. The simulation results are confirmed with direct observations of the microscale droplet interface dynamics as provided by the high resolution holographic measurements. The pressure-interface feedback model accurately predicts the on-source vibration amplitude required to initiate capillary waves, and interfacial oscillation amplitude and frequency. The radiation pressure distribution is likewise confirmed with particle migration observations. Viscous effects on wave attenuation are also studied by comparing experimental and simulated results for a pure water droplet and 90\,wt\%-10\,wt\% glycerol-water solution droplet.
\end{abstract}

\maketitle


\section{\label{sec:level1}Introduction}

High-frequency ultrasound (above $\mathcal{O}(10^{6})\text{ Hz}$) is useful in a variety of applications in micro-fluidic devices such as droplet manipulation, fluid mixing and atomization because of the compatible wavelength for microfluidics. Moreover, the setup for ultrasonic devices are usually simple, and are therefore suitable for constructing small, compact experiments. Nozzle-free atomization methods with high-frequency ultrasonic devices were reported by Ang \textit{et al.}~\cite{ang2015nozzleless} and Collignon \textit{et al.}~\cite{collignon2018improving}. The devices reported in these works are energy efficient, and are able to atomize fluid at high flow rates. A nozzle-free implementation is also drastically simpler: the fluid is atomized into small droplets directly from the transducer surface.

Due to the complexity of the nonlinear interactions between the fluid and the acoustic wave, many of the physical phenomena related to high-frequency acoustically induced atomization are not well understood. At the relatively low forcing frequencies encountered in every day settings, atomization is successfully explained by the classical Faraday instability theory. However, forcing frequencies typically employed in modern acoustofluidics severely violate a fundamental Faraday wave theory assumption: the difference between the excitation frequency and the natural resonant frequency should be much smaller than the excitation frequency~\cite{perlin2000capillary,binks1997nonlinear}. In systems that violate the Faraday conditions, there is a complete absence of any classically predicted originating mechanisms for resonant capillary wave generation, yet such waves are nonetheless found at scales that are visible to the eye.

Acoustically induced capillary waves were first described by Rayleigh in the nineteenth century~\cite{rayleigh1879capillary}. Over the last several decades, researchers have begun employing ultrasound to force the dynamics of the droplets and investigating the effects. With power input that is high enough, capillary waves on the liquid-gas interface lose its stability and small droplets are atomized from capillary wave crests~\cite{lang1962ultrasonic}. Compared to traditional jet nebulizers, ultrasonic nebulizers are more portable, efficient, and easy to use. Ultrasonic nebulizers are widely used in pulmonary drug delivery~\cite{taylor1997ultrasonic}, surface coating~\cite{majumder2010insights}, and many other fields. However, most of these studies focus on the behavior of the droplet excited either with low frequency vibration of a plate~\cite{whitehill2010droplet} or with  modulated acoustic waves using ultrasonic devices. Trinh and Wang modulated the ultrasound to excite the vibration of droplets in a liquid-liquid system~\cite{trinh1982experimental}. Michael Baudoin~\cite{baudoin2012low} modulated the surface acoustic wave (20~MHz) to waves with frequencies lower than 150~Hz. Less energy is required in this case to vibrate and move the droplet relative to the unmodulated signal. In these studies, although ultrasonic waves are used, it is low-frequency resonance interactions with the high-frequency vibrations that generate the oscillation of the droplets while the original high-frequency ultrasound is regarded as a static radiation stress source. The excitation frequencies used in these studies are near the resonant frequencies of the droplets. Blamey~\textit{et al.} \cite{Blamey:uq} studied capillary waves induced with ultrasonic forcing frequency $\mathcal{O}[10^{7}\text{ Hz}]$. Remarkably, they observed capillary waves at the droplet's natural frequency ($\mathcal{O}[10^{2}\text{ Hz}]$) and a \emph{complete absence} of evidence of Faraday instability. The mechanism of energy transfer across these vastly disparate scales was left unresolved.

As for the theoretical study of the droplet vibration excited by acoustic waves, Murray \textit{et al.}~\cite {murray1999droplet} first applied the boundary element method (BEMs) to simulate a droplet's response to acoustic excitation. Lyubimov theoretically studied the oscillatory behavior of a hemispherical droplet on an oscillating substrate~\cite{lyubimov2006behavior}. The model in that work explicitly considers the effect of the pinned contact line. These foregoing numerical studies are limited to forcing scenarios where the ratio of the excitation frequency to droplet resonant frequency is less than ten. Models and observations limited to this regime are insufficient for characterizing droplet vibration in response to ultrasonic acoustic forcing, since the difference between droplet resonant frequency and the forcing frequency is typically many orders of magnitude.

An important assumption made in theoretical studies of the droplet oscillation is that the perturbation is infinitesimal~\cite{strani1984free}. For low frequency ultrasound, since the wavelength is much larger than the droplet's characteristic length, one can assume that the shape of the interface is only infinitesimally distorted when the input power of the ultrasound is small. In this case, a droplet's shape can be expressed as a sum of Legendre polynomials. This approach is infeasible for droplets excited by high frequency 
ultrasonic waves. When the wavelength of the acoustic wave is comparable or even smaller than the radius of the droplets, induced pressure nodes will cause the droplet surface to elevate and statically deform. This static abnormal curvature is observed in Manor's study of a 2~$\mu\ell$ droplet atop a lead zirconate titanate (PZT) thickness polarized disk transducer operating at 2~MHz~\cite{manor2011substrate}. Suryanarayana~\cite{suryanarayana1991effect} studied the effect of the shape change caused by the acoustic radiation stress for levitated droplets. However, in this study, the shape change is regarded as a static deformation and the interaction between fluid shape and pressure distribution is ignored by assuming the deformation is static.

In this letter, we describe a physical model that explains the energy transfer from high-frequency ultrasonic forcing (MHz and beyond) to low-frequency capillary waves on the droplet surface based on interaction between the acoustic radiation force and surface tension.

\begin{figure*}[t!]
    \centering
    \includegraphics[width=0.9\textwidth]{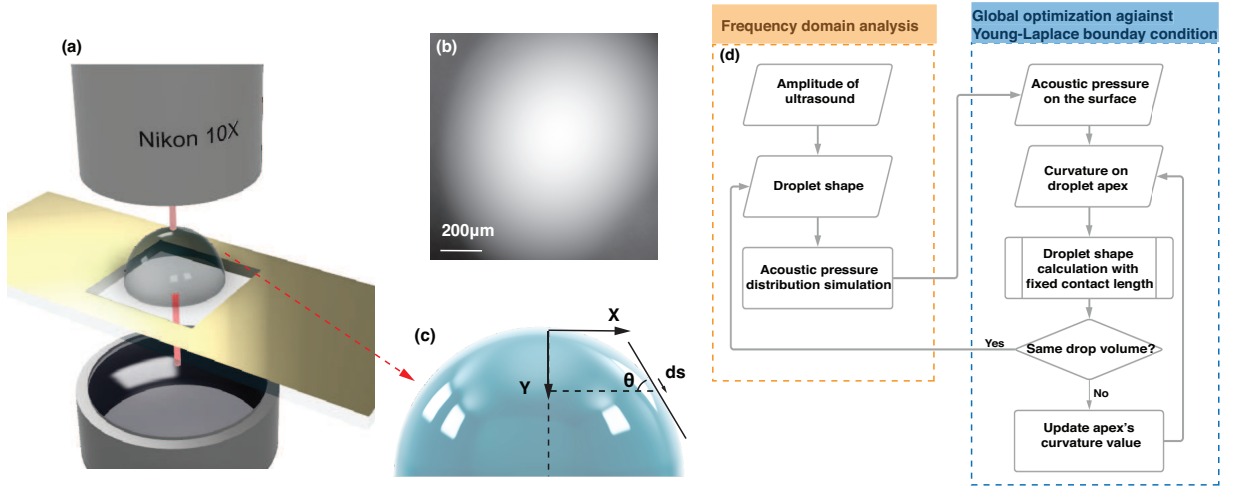}
    \caption{
    (a) Experimental setup with the DHM system and thickness mode device; laser light comes from the laser condenser at the bottom, passes through the droplet sample placed on the top surface of the acoustic device, and then is collected by the lens above. (b) Image containing information on the phase difference at the center of droplet's surface from the DHM system. (c) A droplet placed on the surface of the acoustic device in the transparent window area. The (d) flow chart of the algorithm simulating the shape of the droplet and the acoustic pressure distribution.}
    \label{experiment}
\end{figure*}

\section{Experimental Methods}

The ultrasonic devices were fabricated from $128^{\circ}$ Y-rotated, X-propagating lithium niobate wafers with 500~$\mu$m thickness and mirror-finish polishing on both sides (Roditi, London, UK). On each side of the wafer, a sputter deposition method (Denton Discovery 18, New Jersey, USA) was used to deposit a 400~$nm$ layer of chromium and a layer of gold. These provide electrodes for the thickness mode vibration of the substrate. One 0.5$cm \times$0.5$cm$ transparent area was left without gold deposition at the center of each transducer for the digital holographic microscope (DHM) laser to pass through the media during experiments (Fig.~\ref{experiment}(a)). Thickness-mode vibrations were induced by applying an amplified voltage potential at a frequency matched to the thickness resonance of the device (6.6~MHz for the 500~$\mu m$ thick wafer). A 5-$\mu\ell$ droplet of deionized water was dispensed onto the center of the transparent window using a measuring pipette (2-20 $\mu L$, Thermo Fisher Scientific, USA). This droplet volume is used so that the radius of the droplet is smaller than the capillary length, $l = \sqrt{\gamma/\rho g}$, with $\gamma$ as surface tension and where $\rho$ is the liquid density. For the media used in this study, capillary forces dominate at the droplet surface and the effect of gravity is inconsequential.

The resonant frequency and vibration amplitude per unit volt input for the transducer were characterised with laser Doppler vibrometry (LDV; UHF-120, Polytec, Waldbronn, Germany). Measuring microscale vibrations on the surface of droplets is challenging due to the size and speed of the dynamics under consideration (nanometer amplitude at timescales as small as a few milliseconds). While an LDV is suitable for single-point and scanning measurements of a surface with well-defined periodic vibrations, the DHM (transmission, Lyncee-tec, Zurich, Switzerland) utilizes cutting-edge metrology to characterize interfacial dynamics across an entire region of interest on the liquid-air boundary. The transmission DHM system used in this study generated three-dimensional holographic data by interpreting comparative phase delays between a laser that passed through the dynamic medium and a reference laser traversing an unobstructed path. Although the phase was only unique up to a $2\,\pi$ factor, continuous changes in space and time allow for phase unwrapping of the two-dimensional images to reliably overcome this constraint. When combined with the refractive index of the medium, the unwrapped images provide high-accuracy measurements of time-dependent interfacial displacements. This made the DHM system particularly well-suited for measuring capillary waves on an air-fluid interface. We employed a high-speed camera (FASTCAM NOVA S12, Photron, San Diego, CA USA) integrated with the DHM. In total, the system offers a recording rate of up to 116,000~fps and provided real-time three-dimensional surface structure data with high spatial resolution (nanometer scale accuracy in the normal vertical $y$-direction and micron scale accuracy in the transverse $xz$-plane).

\section{Physical model}

The acoustic impedance can be calculated by the density of the media and speed of sound $Z = \rho c$. The reflection coefficient, which describes how much portion of the acoustic power is reflected at the interface, can be expressed as $r = {\frac{Z_l-Z_g}{Z_l+Z_g}}^2$, where $Z_l$ and $Z_g$ are the acoustic impedances of liquid and gas. With water and air used in this example, more than 99\% of the acoustic waves carried in water are reflected from the air-water interface, implying that most of the vibrational energy will be conserved in the droplet. Since the attenuation length of the acoustic wave can be estimated by Stokes' law~\cite{lighthill1978acoustic}: $\alpha_l^{-1}=\frac{\rho v^3}{4 \pi^2 f^2 (\frac{4}{3}\mu+\mu_b)}$\, where $\rho$ is the density of the liquid, $v$ is sound velocity, $f$ is the frequency, and $\mu$ and $\mu_b$ are the dynamic and bulk viscosity, respectively, the attenuation length of 1-MHz-order acoustic waves are more than a meter. The acoustic waves are expected to be reflected multiple times and form compressed and rarefied regions. The acoustic pressure pattern formed in the droplet would thus affect the fluid surface in a complicated way. Mass and momentum conservation equations are used in the analysis~\cite{nyborg1965acoustic,riaud2017influence} to account for this complexity;
\begin{subequations}
    \begin{align}
        \begin{split}
            \frac{\partial\rho}{\partial t}+\nabla\cdot(\rho u)=0
        \end{split}\\
        \begin{split}
            \rho\frac{\partial u}{\partial t}+\rho(u\cdot\nabla)u=-\nabla p+\mu\nabla^2u+(\mu_B+\frac{\mu}{3})\nabla\nabla\cdot u)
        \end{split}
    \end{align}
    \label{NS}
\end{subequations}



Where $\rho$ is the fluid density, $u$ is the fluid velocity, $P$ stands for the fluid pressure, $\mu$ and $\mu_B$ represent shear viscosity and bulk viscosity respectively. Since the vibrational velocity from the ultrasonic device required to initiate the capillary wave in our study is small, the so-called 'slow streaming' assumption \cite{Friend:2011ss} can be used in the analysis and the physical quantities in the equation \ref{NS}(a) and \ref{NS}(2) can be decomposed into three contributions~\cite{hunt1955notes,nyborg1965acoustic} as:
\begin{subequations}
    \begin{numcases}{}
      u = u_0 +\epsilon u_1 +\epsilon^2u_2 +\mathcal{O}[\epsilon^3]\\
      p = p_0 +\epsilon p_1 +\epsilon^2p_2 +\mathcal{O}[\epsilon^3]\\
      \rho = \rho_0 +\epsilon \rho_1 +\epsilon^2\rho_2 +\mathcal{O}[\epsilon^3]
    \end{numcases} 
    \label{decompose}
\end{subequations}

$u_0$, $p_0$ and $\rho_0$ are hydrostatic terms and those with subscript 1 and 2 refers to first and second order perturbations. $\epsilon$ is Mach number defined as ratio of fluid velocity and speed of sound ($\epsilon = u_1/c_0$). Based on the fact of small velocity, $\epsilon\ll 1$. Taking expansion \ref{decompose} (a-c) into equation \ref{NS}(a) and \ref{NS}(b) and grouping in terms of $\epsilon$, the equations can be separated into three parts: the zeroth, first, and second order components of the acoustic perturbation. The expression of the first-order acoustic perturbation represents the behavior of the linear acoustic waves in the fluid. Since the size of the droplet is small, the hydrodynamic Reynolds number in this case would be a small value and the equations can be further simplified~\cite{zarembo1971acoustic} as:
\begin{subequations}
    \begin{align}
        \begin{split}
            \frac{\partial \rho_1}{\partial t}+\rho_0(\nabla\cdot u_1) &= 0
        \end{split}\\
        \begin{split}
            \rho_0\frac{\partial u_1}{\partial t} &= -\nabla p_1
        \end{split}
    \end{align}
    \label{NS2}
\end{subequations}



Together with the linear equation of state $p_1 = c_0^2\rho_1$, these equations can be used to describe the acoustic wave in the fluid with small Mach and Reynolds numbers. We solve for the radiation pressure using the linear pressure wave equations above. The acoustics are modified to include reflection on the interfaces and attenuation along the path of propagation. We use the finite element method in the frequency domain to obtain the pressure distribution (COMSOL Multiphysics, COMSOL Inc., Burlington, MA USA). The impedance boundary condition is used to simulate the reflection of the acoustic wave on the fluid-air interface. The acoustic wave pressure is assumed to decay exponentially with distance when travelling in the fluid. The attenuation factors are calculated based on the properties of the fluid and the acoustic waves \cite{takamura2012physical}.

Manor \textit{et al.}~\cite{manor2011substrate} have reported that the acoustic radiation pressure on an air-water interface generated with ultrasonic actuator working at high frequency could cause the droplet to deform. The pressure jump at the interface and surface curvature (\emph{i.e.}, shape) are related according to the Young-Laplace boundary condition. We use an axisymmetric droplet shape analysis method~\cite{del1997axisymmetric} to numerically minimize the droplet shape error subject to a constant volume constraint $V_0$ and fixed contact length constraint $l_0$. Thus, the classical Laplace equation can be expressed as function of arc length $s$ on the interface and tangential angle $\theta$:
\begin{equation}
    \frac{d\theta}{ds} = 2b+cz-\frac{\sin\theta}{x}+\frac{P_a}{\gamma},
    \label{laplace}
\end{equation}
with 
\begin{subequations}
    \begin{align}
      {dx}/{ds} &= \cos\theta\\
      {dy}/{ds} &= \sin\theta\\
      {dV}/{ds} &= \pi x^2\sin\theta
    \end{align} 
    \label{three}
\end{subequations}

The problem is simplified into a two-dimensional case based on the axisymmetric assumption; $x$, $y$ and $V$ represent the position and the differential volume at the corresponding position. The acoustic pressure $p_a$ on the interface, as the result from previous simulation, is extracted to calculate the surface shape of the fluid. We name this simulation process the \emph{pressure-interface feedback model} since it mimics the feedback interplay process between the acoustic pressure distribution and the shape of the droplet's interface. In  eqn.~(\ref{laplace}), $b$ is the curvature at the apex, which is treated as an additional variable to solve the equation with a Neumann boundary condition (${d\theta}/{ds}=b$ at $s=0$). It can be seen from eq.~(\ref{laplace}) that only the change of angle over different locations $s$ is updated with this numerical method, which means that though the drop's surface shape can be calculated, the step size $ds$ is required to determine the length of the curve. In this paper, a fixed step size is used to calculate the drop shape. To increase accuracy, a tiny fixed step size ($10^{-7}$~m) compared to the drop's dimension was used. A small step size requires a finer grid and a larger number of points when extracting pressure data from the simulation result. With an artificially defined step size, it remains difficult to obtain an exact solution of the drop shape, though the overall shape should be correct. To overcome this problem, the shape of the droplet is then rescaled so that the contact length matches with $l_0$. Another challenge is that the curvature $b$ is unknown without measurement with specific devices. A shooting method is therefore applied to generate results from eqns.~\eqref{laplace} and \eqref{three}. Each time a different value of $b$ is found in this process and used to solve the differential equation \ref{laplace} until the volume conservation condition is satisfied ($\sum_i dV_i = V_0$). The pressure-interface feedback model is implemented with two nested loops as shown in Fig.~\ref{experiment} (b) with the outer loop seeking values of the curvature on the droplet's apex via the shooting method, and the inner loop solving the differential equation in a stepwise fashion, terminating when the last pressure data point is reached. After calculating the surface shape of the droplet, the interface is then updated and imported back into the finite element analysis to determine the acoustic pressure distribution for the next quasi-static state calculation.




\begin{figure*}[t!]
    \centering
    \includegraphics[width=0.9\textwidth]{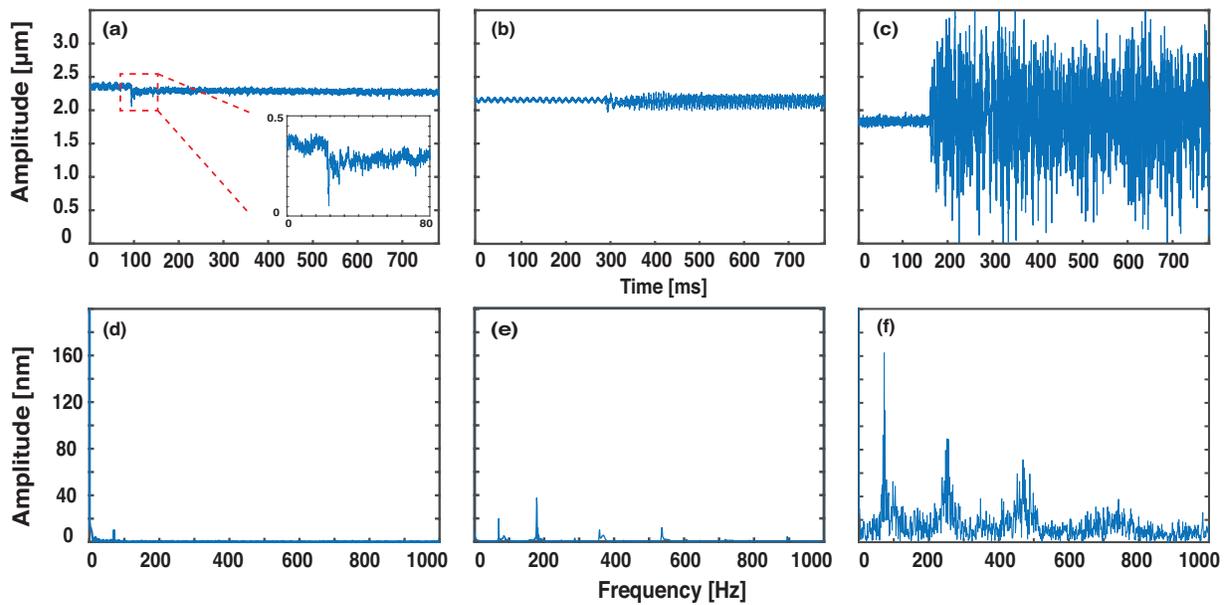}
    \caption{Vibration pattern of the droplet collected with the DHM system before and after the power is applied to the acoustic devices, the amplitude of the acoustic waves are 1.1~nm (a), 1.6~nm (b) and 2.3~nm (c) respectively.(d)-(f) are the FFT of the results in (a)-(c) separately after the power is applied to the acoustic devices }
    \label{f2}
\end{figure*}
We next seek to correlate the interframe time scale $\Delta t$ to an approximate expression derived from nondimensional analysis. The equation $\Delta x = 1/2a\Delta t^2$ is used since, initially, the vibrational velocity on the droplet surface is zero. The expression $\Delta x$ represents the displacement of the droplet's  interface at its apex, and $a$ is the acceleration.  Thus the inter-frame time can be estimated to be $\Delta t = \sqrt{2\Delta x/a}$. Acceleration on the interface is caused by the acoustic pressure $P_a$, and if we treat the points on the interface as differential surface areas $A$, the relationship between the acoustic pressure and the acceleration can be defined as $p_a A = ma$. Thus $a = \frac{P_aA}{\rho V}=\frac{P_a}{\rho \Delta x}$. Taking the equation of acceleration into the expression of the inter-frame time, we obtain $\Delta t = \sqrt{\frac{2\rho}{P_a}}\Delta x$. Based on the simulation result, the inter-frame time ranges from $10^{-4}$ to $10^{-3}$~s, which is much larger than the period of the excitation signal ($10^{-7}$~s).


\section{Results and Discussion}

Experiments revealed three essential dynamical regimes: static shape change, steady vibration, and nonlinear vibration, as shown in Fig.~\ref{f2}~(a)-(c), respectively. Care was taken to isolate the system from ambient perturbations such as vibration and localized air currents. The residual vibration with an amplitude of around 60~nm on the droplet surface is due to the high-speed camera fan vibrating the observation system (the high-speed camera is rigidly fixed to the observation tray). 
We then studied the effect of different on-source vibrational amplitudes on the oscillation of the droplet surface and we controlled the amplitude by tuning the input power to the ultrasonic devices. The amplitudes of the thickness-mode vibrations on the transducer surface were detected with LDV. The noted dynamical regimes correspond to the on-source vibration amplitude. When the on-source vibrational amplitude is small ($\leq 1.5$~nm), a sudden change of the droplet height is observed at the instant acoustic excitation is applied (Fig.~\ref{f2} (a)). This occurs due to a sudden change in the pressure at the interface resulting from acoustic radiation forces. In the static shape change mode, an increase in input power to the transducer corresponds to an increase in deformation, and the amplitude of the droplet surface oscillation before and after the droplet's shape is unchanged. Evidently the high-frequency acoustic wave is not directly interacting with the low-frequency oscillation on the droplet surface. This suggests the existence of another mechanism facilitating energy transfer from the ultrasonic to the capillary wave's wavenumbers. The natural oscillation of the droplet surface can be predicted with Rayleigh's equation,
\begin{equation}
    f = \frac{1}{2\pi}\sqrt{\frac{l(l+1)(l+3)\gamma}{\rho R^3}},
\end{equation}
where $R$ is the radius of the droplet, $\gamma$ is the surface tension, and $l=1,2,3,...$ is the mode number. In this case, the first natural frequency is 80~Hz, which agrees with the stationary frequency transform in Fig.~\ref{f2} (d). The results in Fig.~\ref{f2}~(b) are generated with a 1.5~nm input. The oscillation amplitude on the droplet surface is distinct from the ambient fan perturbation in amplitude and frequency. Sudden shape change can still be observed when input signal is initiated. Following the shape change, the droplet exhibits linear, stable vibrations. Frequency peaks associated with these vibrations are plotted in Fig.~\ref{f2} (e). These are peaks located at 178, 357 and 537~Hz. No super-harmonic modes for the natural frequency are observed. To clarify how the energy is transferred from the ultrasonic device's vibrations to capillary waves, we conducted particle image velocimetry (PIV) experiments with high-speed camera (FASTCAM MINI, Photron, Japan). Acoustic streaming can be generated by nonlinear interaction between acoustic wave and fluid \cite{lighthill1978acoustic}. This phenomenon is commonly seen when the frequency of the ultrasound is high since higher proportion of acoustic energy will be attenuated into the fluid to drive the flow \cite{dentry2014frequency}. However, there is no flow observed with the same amplitude of the vibrations from the source that is able to drive the steady vibration on the droplet surface. Thus the observed capillary waves are not a result of any flow behavior.

As on-source vibration amplitude is increased, nonlinearity plays a larger role in the capillary wave dynamics. Evidence of nonlinearity is obsserved in Fig.~\ref{f2} (c) and (f). In Fig.~\ref{f2} (c), the wave pattern is nonuniform and no obvious period of oscillation can be directly observed. The peaks in the frequency space are broadened and this is due to non-resonance interaction between waves with different frequency (Fig.~\ref{f2} (f)). These interactions generate waves with new wavelengths and frequencies. With a low-level of nonlinearity in a finite domain, a wave that is not congruous with the resonance conditions of the droplet will vanish while newly generated waves that are located within a spectral range determined by the nonlinear broadening of the dispersion relation will remain~\cite{berhanu2018turbulence,kartashova2010nonlinear}.

 We then confirmed the normal axis DHM measurements of the rapid, transient initial shape change with direct high speed transverse profile imaging recorded by high speed camera. Since the initial shape change of the droplet is in the submicron to micron scale, we used a high-speed camera with a 5X objective lens (M Plan Apo 5x objective, Mitutoyo, Japan) to observe the droplet from the side and capture the height difference before and after the power was applied to the ultrasonic device. 
 We then binarized the images that were produced from the camera to then calculate the height change at the droplet apex.
\begin{figure}[!b]
    \includegraphics[width=0.5\textwidth]{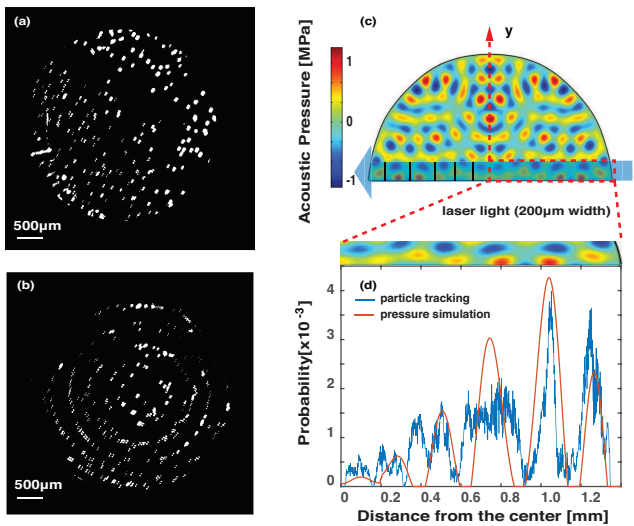}
    \caption{Two pictures taken in the particle tracking experiment process (a) before and (b) after the acoustic device was turned on. A ring-shaped pattern was formed by acoustic streaming-driven recirculation to form nodes corresponding with the location of the high-pressure regions in the droplet; this is comparable to (c) the two-dimensional (side view) acoustic pressure distribution in the water droplet, taking advantage of the axisymmetric nature of the droplet. The particles' position from the particle tracking experiments (d, blue line) favorably compare to the peaks in pressure from the simulation (d, red line).}
    \label{fp}
\end{figure}

In order to resolve the pressure distribution,  we tracked the migration of a homogeneous dispersion of fluorescent polystyrene particles (3~$\mu$m Fluoresbrite YG Microspheres at a concentration of $4.99 \times 10^5$~particles/m$\ell$; excitation and emission maximum wavelengths at 485 and 441~nm, respectively, Polysciences, Warrington, PA, USA) using high-speed imaging. The size was selected to be much smaller than the wavelength of the progressive acoustic wave in order to mitigate the  influence of direct acoustic radiation forcing. This ensures that the particles will only migrate due to local acoustic streaming that will deliver the particles to the high pressure regions which are the quiescent nodes amid the recirculating flow adjacent the substrate surface. 
We illuminated the particles with a blue laser sheet generator (488~nm wavelength). To decrease the background light intensity, a low-pass optical filter (<450~nm longpass filter FEL0450, ThorLabs, Newton, NJ USA) was placed in the camera light path to block the excitation light, leaving only the fluorescence signal to pass to the camera. The thickness of the laser sheet is 200~$\mu$m, only the particles in this height range will be illuminated. We set up the laser to pass through the bottom of the droplet so more particles can be visualized. Figure~\ref{fp} (a), (b) shows the distribution of the particles before and after acoustic excitation. The particles are uniformly distributed in the droplet before the acoustic wave is generated and migrate to well-defined positions forming a ring-like pattern during excitation.
\begin{figure*}[!t]
    \centering
    \includegraphics[width=0.8\textwidth]{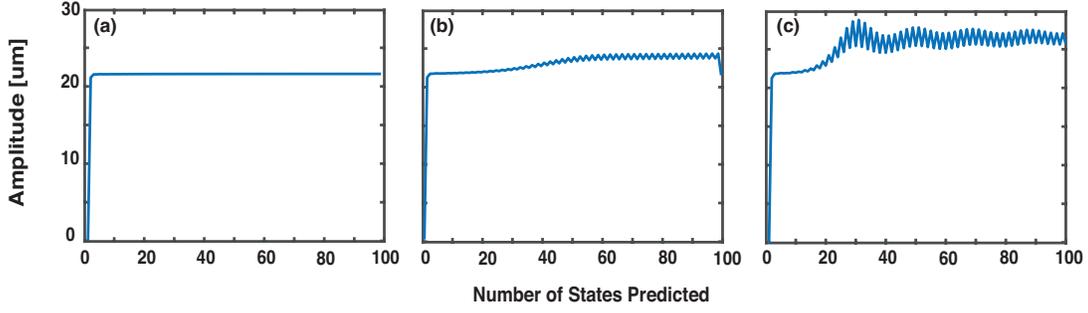}
    \caption{The simulation results of the motion of the apex of a droplet's fluid interface excited by thickness-mode vibrations: (a) 1.1~nm amplitude vibrations are insufficient to induce vibration after the initial, nearly static height change. (b) A 1.5~nm vibration was required in the simulation for the droplet's interface to exhibit steady vibrations.  Increasing the vibration to (c) 1.9~nm on the droplet surface amplifies the motion and higher-order motions are also apparent. These vibration amplitude-driven thresholds from a static shape to steady vibration and nonlinear vibration are consistent with the experimental results in Fig.~\ref{f2}.}
    \label{fs}
\end{figure*}

 The results of the acoustic pressure simulations are shown in Fig.~\ref{fp} (c). The simulations take into account wave reflection and attenuation. The complex distribution of positive and negative pressure nodes is caused by the interaction of acoustic waves with the interface as they are reflected multiple times within the droplet. The pressure wave interactions lead to local pressure nodes. Within a stable oscillating pressure distribution, particles are driven from positive pressure nodes to the closest positions with negative acoustic pressure. The results of our simulation are confirmed with experimental particle migration measurements. The numbers of particles with different distance are counted and the histogram is normalized to the blue curve in Fig.~\ref{fp} (d). Each point on the curve represents the probability that a particle will be located at a specific distance from the droplet center after migration. To compare the experimental data to the simulated position of the negative pressure nodes, we take the average of the pressure simulated in different layers along y axis at the bottom of the droplet (blue area shown in Fig.~\ref{fp}(c)). Since the particles migrate toward the closest negative nodes, the probability associated with a particle migrating to a give position is proportional to: (i) the pressure, and (ii) the number of particles in the regions. We divide the illuminated area into several regions according to the midpoints between any two neighboring negative pressure nodes (the way of how these regions are separated are shown with the black lines through the midpoints in Fig.~\ref{fp}(c)). We calculate the ratio of particle counts in different regions by comparing the area of the annular regions. Particles within a certain region are assumed to migrate to the local negative pressure node. The red curve in Fig.~\ref{fp}(d) represents the normalized probability corresponding to migrated particle positions based on the simulated pressure results. The data collected from the particle tracking experiments find good agreement with magnitude, number, and location of the pressure nodes predicted by the pressure-interface feedback model. Since there is no significant net acoustic streaming flow within the droplet, these results provide strong evidence for the existence of the spatially localized stable pressure distribution. It can be seen here that with high frequency ultrasound, the acoustic waves' wavelength is on the order of, smaller than, the size of the droplet. When properly accounted for, the effects of reflection and attenuation of the acoustic waves and their interactions serve to redistribute pressure within the droplet in a manner that is highly consistent with our observations. This demonstrates a clear, intuitive mechanism for the noted energy transfer across wavenumbers spanning many orders of magnitude. And this is clearly very different from the mechanism(s) proposed by the classical theory.


In order to analyze the dynamic droplet shape change induced by acoustic pressure feedback, we constructed a pressure-interface model by extracting the simulated pressure data from the surface of the droplet and utilized the data within a modified Young-Laplace boundary condition (eqn.~\eqref{laplace}). Here, the surface tension balances the acoustically-driven dynamic pressure jump by inducing local curvature. The direction of the change is determined by the sign of the local pressure change. At each step in the the simulation, the shape that is deduced by minimizing the curvature against the Young-Laplace boundary is then utilized to compute an updated pressure distribution. This update is then used with the Young-Laplace condition to update the droplet shape. Iterating accordingly, we obtain a time series of states of the droplet shape and pressure distribution. Figure \ref{fs} (a) shows the simulated case for a small on-source vibrational amplitude: the transducer amplitude is 1.1~nm. The droplet experiences a nearly instantaneous height change when the input is switched on and stabilizes with the new shape and no further changes appear. This corresponds to experimental observations of the static mode where capillary waves are not generated. However, with a slightly greater input amplitude of 1.5~nm, capillary waves are generated on the droplet surface. In the simulation results shown in Fig.~\ref{fs} (b), one observes that at this amplitude, the droplet apex vibrations corresponds to a capillary wave with an amplitude of 200~nm. This prediction is also consistent with the experimental results. As the forcing amplitude is further increased to 1.9~nm, the amplitude of the resulting capillary waves increases to 1~$\mu$m and nonlinear vibration patterns emerge, as shown in Fig.~\ref{fs} (c).


\begin{figure}[ht]
    \centering
    \includegraphics[width=0.5\textwidth]{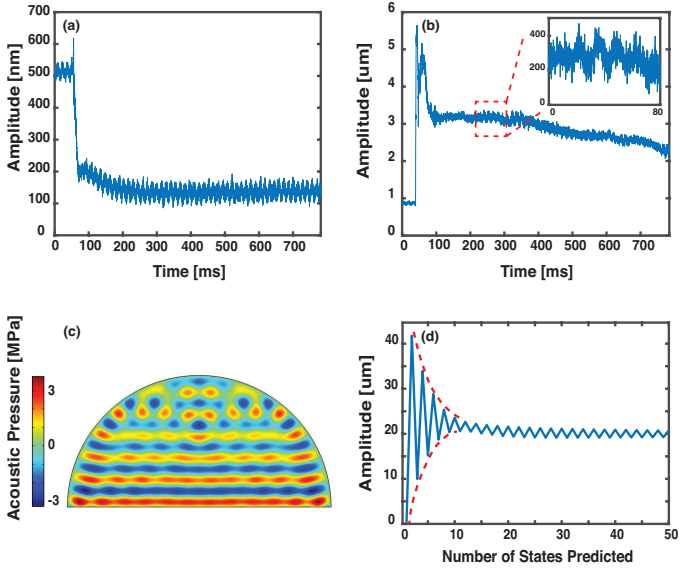}
    \caption{ Experimental results of 90\%-10\% glycerol-water solution droplet excited with acoustic waves in static mode (a) and steady vibration mode (b) (3.9~nm amplitude). (c) shows the laminar pattern of the pressure distribution in the 90\%-10\% glycerol-water solution droplet. (d) The simulation results of the vibration generated with 3.9~nm acoustic waves at the apex of the droplet} 
    \label{f5}
\end{figure}
The complexity of the pressure distribution is due not only to wave reflections, small attenuation is also an important factor. The attenuation factor for acoustic waves is~\cite{morse1986theoretical} $2(\frac{\alpha V}{\omega})^2=\frac{1}{\sqrt{1+\omega^2\tau^2}}-\frac{1}{1+\omega^2\tau^2}$, where $\alpha$ is the attenuation coefficient, $V$ is the sound velocity, $\omega$ is the angular frequency, and $\tau$ is the relaxation time. Here the relaxation time is given by $\tau = \frac{4\mu/3+\mu_b}{\rho v^2}$. For water droplets, the dynamic viscosity $\mu$ and bulk viscosity $\mu_b$ are $0.89$~mPa$\cdot$s and $0.2$~mPa$\cdot$s, respectively. For an acoustic wave at 6.6~MHz, the attenuation distance ($1/\alpha$) is therefore 0.034~m, so the acoustic waves are reflected multiple times until fully attenuated within millimeter sized water droplet, like those considered here. To study attenuation effects on capillary wave formation, we conducted experiments and simulations for a 90\%-10\% glycerol-water solution. Glycerol is used since it has a similar density (1260~$kg/m^3$) and surface tension (63.4~$mN/m$) to water, but a substantially higher attenuation ($1/\alpha = 2.8\times10^{-4}$~$m$). This allows us to isolate the effect of attenuation on capillary wave formation. The results for the solution are similar to those for water. With small vibrational amplitude, only sudden height jumps are observed (Fig\ref{f5} (a)) and no capillary wave formation. The vibration amplitudes on the interface before and after the droplet height jumps are same, which means that droplet only changes from one steady state to another in the process. Figure \ref{f5} (b) shows the vibration of the droplet excited by input with 3.9~nm amplitude. The details of the vibrations can be seen in the inset. The vibration at droplet apex itself is of the linear sine wave form: with 3.9~nm amplitude input, capillary waves are generated on the surface and the droplet reaches a steady vibration mode. Compared to the vibration of the water droplet, the solution droplet took significantly longer to reach the steady vibration state after the sudden droplet shape change. Simulations were conducted with the same parameters used in the experiments and the results for the 3.9~nm input amplitude are shown in Fig.~\ref{f5} (c), (d). Figure \ref{f5} shows the acoustic pressure distribution in the droplet. A laminar pressure distribution is observed with nodal formation near the top portion of the droplet. The input amplitude threshold for capillary wave generation is confirmed with experiment, as shown in Fig.~\ref{f5} (b), further validating the model. A rapid capillary wave amplitude decay is observed in both the simulation and the experiment. The ratio between the decay time to the period of the steady vibration is roughly five in both cases, as shown in Fig.~\ref{f5} (b)(d).

\section{Conclusions}

A new method to observe the onset and growth of capillary wave motion on fluid interfaces from high frequency acoustic waves has been provided using high-speed digital holographic microscopy. The results produced from this method are compared to a new approach to the solution of capillary wave dynamics through the use of a hybrid solution method. This method employs a two-step process, first producing the pressure distribution on the fluid interface from the relatively fast acoustic standing wave distribution in the acoustic cavity formed by the droplet. This step is followed by a computation of the new shape of the fluid interface that would arise as a consequence of the new pressure distribution taking into account the acoustic pressure variation at the interface. The necessary increment in time between steps was determined from a simple nondimensional analysis. Remarkably, the correlation was good between the computational results produced using this method and the experimental observations. Further refinements of this method is likely to produce additional insight into the complex phenomena of capillary wave generation.

\section{Acknowledgments}
The work presented here was generously supported by a research grant from the W.M.\ Keck Foundation to J.\ Friend. The authors are also grateful for the substantial technical support by Yves Emery and Tristan Coloumb at Lyncee-tec, and Eric Lawrence, Mario Pineda, Michael Frech, and Jochen Schell among Polytec’s staff in Irvine, CA and Waldbronn, Germany. Fabrication was performed in part at the San Diego Nanotechnology Infrastructure (SDNI) of UCSD, a member of the National Nanotechnology Coordinated Infrastructure, which is supported by the National Science Foundation (Grant ECCS--1542148).

%


\end{document}